\documentclass[prc,superscriptaddress,showpacs,preprintnumbers,amsmath,amssymb,floatfix]{revtex4}

\usepackage{graphicx}
\usepackage{dcolumn}
\usepackage{bm}
\usepackage{longtable}

\begin{document}

\title{ Nuclear Structure of the even-even Argon isotopes with a focus
  on magnetic moments}

\author{S.J.Q. Robinson}

\affiliation{
Department of Physics,
Millsaps College,
Jackson, MS 39210,
}

\author{Y.Y. Sharon}
\affiliation{
Department of Physics and Astronomy,
Rutgers University,
Piscataway NJ 08854
}

\author{L. Zamick}

\affiliation{
Department of Physics and Astronomy,
Rutgers University,
Piscataway NJ 08854
}

\date{\today}

\begin{abstract}

We study the role of configuration mixing in the heavier even-even
isotopes of Argon.  We begin by limiting the configurations of the
even-even Ar isotopes to $(d_{3/2}^2)_{\pi}$ $(f_{7/2}^n)_{\nu}$.
There, due to the particular location in this shell model space of
$^{40}$Ar and $^{44}$Ar, we find that the spectra, B(E2)'s and
magnetic moments of these two nuclei are identical. Any deviation from
this equality is direct evidence of configuration mixing. In a larger
shell model space there are significant differences between these two
nuclei, with $^{44}$Ar being more collective.  We also consider other
even-even isotopes of Argon and study how their nuclear structure
effects evolve with N.  We compare in the full 0$\hbar \omega$ space
$(sd)_{\pi}$ $(fp)_{\nu}$ the results of calculations with the WBT
interaction and with the newer SDPF, denoted SDPF-U, interaction.

\end{abstract}

\maketitle

\section{Introduction}

Previously members of our theoretical group were involved in
collaborations with experimentalists from Rutgers and Bonn studying
the properties of $^{36,38,40}$Ar as well as $^{32,34,36}$S
~\cite{args1,args2,args3,args4,args6,speid08} with the focus
being the magnetic dipole g factors of the 2$^{+}_1$ states.  We also
note the results for $^{38,40}$S obtained in experiments carried
out by Davies et al. and by Stuchbery et al.~\cite{stuch1,stuch2}.

In this work we accomplish two things.  The first is that we observe
that due to the half filled nature of the d$_{3/2}$ proton sub-shell
in the results of calculations in the simplest single j shell picture,
$(d_{3/2}^2)_{\pi}$ $(f_{7/2}^n)_{\nu}$, for the even-even Argon
isotopes beyond $^{36}$Ar, certain symmetries emerge. These symmetries
serve as a basis for comparison with large space calculations for
which these symmetries no longer hold.  Secondly, we extend the
calculations of the properties of the even Ar nuclei to the more
neutron rich Ar isotopes for which the 2$^{+}_1$ g factors have not
yet been measured. These are $^{42,44,46}$Ar. In particular, we note
interesting and important divergences in the results obtained
with two different interactions in this region; the WBT
interaction~\cite{wbt} and the updated SDPF interaction denoted SDPF-U
~\cite{oldsdpf,newsdpf}.  The shell model calculations using the WBT
interaction were carried out with the OXBASH shell model
code~\cite{oxbash} and those using the SDPF-U interaction with the
ANTOINE shell model code~\cite{antoine}.  In all the calculations, the
effective charges $e_p=1.5$ and $e_n=0.5$ were used as well as the
free nucleon g factors.

We show in Table ~\ref{tab:gfac}, as a summary, the relevant
previously obtained experimental and calculated g factors of 2$^+_1$
states, data which are contained in the work of Speidel et al. on
$^{36}$S~\cite{speid08}. There are several points of interest. For
example, the g factors of the N=Z nuclei $^{32}$S and $^{36}$Ar are
purely isoscalar and are very close to 0.5, as was discussed before in
papers with Speidel.

\begin{table}
\begin{center}
\caption{Experimental and calculated g($2^+_1$) factors}
\label{tab:gfac}
\begin{tabular}{ccccc}
& Z  &  N       &    experimental g factor & calculated g factor \\
Sulfur & Z=16 & 16 & +0.45 (7) & +0.501\\
   &  & 18 & +0.50(8) & +0.491\\
   &  & 20 & +1.3(5)  & +1.16 \\
   &  & 22 & +0.13(5)$^a$ & -0.055\\
   &  & 24 & -0.01(6)$^a$ & +0.079\\
Argon & Z=18 & 18 & +0.52 (18) & +0.488\\
   &  & 20 & +0.24(12) & +0.309 \\
   &  & 22 & -0.02(2) & -0.195\\
a ~\cite{stuch1}
\end{tabular}
\end{center}
\end{table}

In $^{38}$Ar we have a closed d$_{3/2}$ neutron shell in the above
small space so the g factor is due to the d$_{3/2}$ protons. In this
limit we have g$(2^+_1) ^{38}$Ar = g$(\frac{3}{2}^+_ 1)^{39}$K. The
measured values are +0.24(12) and +0.2610, respectively. The error bars
on $^{38}$Ar are large but the results are thus far not inconsistent.
The measured value for $^{39}$K differs considerably from the bare
value g=+0.083 for a $d_{3/2}$ proton.

In ~\cite{speid08} the comment was made that when one looks at the
g(2$^+_1$) data there is a shell break (at N=20) for the S isotopes but
not for the Ar isotopes. However this can be easily understood.

For $^{36}$S (with N=20) the basic configuration is a s$_{1/2}$ proton
and a d$_{3/2}$ proton coupled to J=2$^+$. This is a stretched
configuration with a g factor 0.5[$\mu s_{(1/2)}+\mu d_{(3/2)}$]. The
answer is dominated by the large s$_{1/2}$ magnetic moment. In the
heavier S isotopes one puts neutrons into the f$_{7/2}$ shell, where
their g factor is negative, and thus a big drop in g(2$^+_1$) is
expected and indeed is observed.

For $^{38}$Ar (with N=20) the basic configuration is d$_{3/2}^2$ protons
with a closed s$_{1/2}$ proton shell. This has a much smaller g factor
than the s$_{1/2}$ d$_{3/2}$ proton configuration.  So when one goes
to the heavier Ar isotopes by adding f$_{7/2}$ neutrons the drop is
not so dramatic.

\section{Small space Equalities}

In order to see the evolution of configuration mixing we calculate the
nuclear structure of several even-even isotopes of Argon in the small
model space $(d_{3/2}^2)_{\pi}$ $(f_{7/2}^n)_{\nu}$ with the WBT
interaction with the A dependence removed.  These results comprise a
part of Tables~\ref{tab:ee},~\ref{tab:g} and~\ref{tab:be2}.

The most striking feature of these results is that one gets identical
results for many properties of $^{44}$Ar and $^{40}$Ar, including the
spectra, B(E2)'s and g factors, as long as one uses the same interaction
for all these nuclei.

Why the equality? It arises from the fact that the two d$_{3/2}$
protons are in mid-shell. In $^{40}$Ar we have two valence f$_{7/2}$
neutrons while in $^{44}$Ar we have two f$_{7/2}$ neutron holes. In
general the particle-hole interaction is not the same as the
particle-particle interaction. But the two d$_{3/2}$ protons are at
mid-shell and so can also be regarded as two d$_{3/2}$ proton
holes. The hole-hole interaction(up to a constant) is equal to the
particle-particle interaction.  This basically proves the relationship
between $^{40}$Ar and $^{44}$Ar in the small space.  As a beautiful
consequence of this relationship, any divergence between these two
nuclei in the $0 \hbar \omega$ larger-space calculations, and more
importantly, in experimental observations, is a direct result of
configuration mixing.

\section{Evolution in the WBT calculations from the small $(d_{3/2}^2)_{\pi}$ $(f_{7/2}^n)_{\nu}$ space to the large $0 \hbar \omega$ $(sd)^2_{\pi}(fp)^n_{\nu}$ space}

a. \underline{The 2$^+_1$ excitation energy in the small and large spaces}\\

   In Table ~\ref{tab:ee} we show the experimental~\cite{exp1,exp2}
   and calculated excitation energies in both the small and large
   spaces for the Ar isotopes using the WBT~\cite{wbt} interaction and
   the OXBASH~\cite{oxbash} shell model code. We also include in that
   table the results of calculations in the larger space with the
   SPDF-U ~\cite{newsdpf}~\cite{oldsdpf} interaction using the ANTOINE
   shell model code ~\cite{antoine}.

In the present section we will compare the small space WBT results
with the large space "0$\hbar \omega$" results using the same
interaction.  By "0$\hbar \omega$" we mean that the protons are
restricted to the full s-d shell and the neutrons to the full f-p
shell. In the next section, we will compare the WBT results with the
results from the more recent SDPF-U interaction.

In the small space the calculated energies of the 2$^{+}_1$ states of
$^{38, 40,42,44,46}$Ar are seen from Table ~\ref{tab:ee} to be
symmetric about $^{42}$Ar. The big increase for $^{46}$Ar is due to
the fact that the eight f$_{7/2}$ neutrons form a closed shell.  The
excitation energy of the 2$^{+}_1$ state of $^{46}$Ar (for a fixed A
interaction) is the same as for $^{38}$Ar.

However, the behavior in the large space is different. There is a
steady decrease in the calculated excitation energy of the 2$^+_1$
state from 2.010 to 1.143 MeV as one goes from $^{38}$Ar to
$^{46}$Ar. Overall there is fairly good agreement between the
calculated large space results and the experimental values except for
$^{46}$Ar. (See Table~\ref{tab:ee}.)

We thus see a problem in understanding $^{46}$Ar, one which we will
address in a later section. The small space calculation yields a rise
in the 2$^{+}_1$ energy of $^{46}$Ar relative to $^{44}$Ar, in
agreement with experiment, but of too large a magnitude. There is no
such rise in results of the large space calculations.  On the other
hand, the observed excitation energy in $^{38}$Ar (2.167 MeV) is
larger than the corresponding energy in $^{46}$Ar (1.550 MeV) --in the
small space they would be the same, at 2.118 MeV.\\

b. \underline{The g factors in the small and large spaces}\\
 
The g  factors for all the  states of interest  were calculated, using
the   free   nucleon  g   factors,   and   are   available  in   Table
~\ref{tab:g}.  We focus  on  the g(2$^+_1$)  values.  For the  reasons
previously  noted,  in  the  small  space the  g(2$^+_1$)  values  for
$^{40}$Ar and  $^{44}$Ar are the  same (-0.441). For $^{46}$Ar  in the
small space, the f$_{7/2}$ neutron shell is filled, and its g(2$^+_1$)
factor is the same (g(2$^+_1$)=+0.083) as that of $^{38}$Ar (being due
to the  d$_{3/2}^2$ proton configuration). The other  Ar isotopes have
negative  g factors  in the  small space  calculations because  of the
contributions of the open-shell f$_{7/2}$ neutrons.

The above small space g(2$^+_1$) numbers change considerably when the
large space is considered. Note that for $^{38}$Ar the change in the
g(2$^+_1$) values from the small space (g(2$^+_1$)=+0.083) to the
large space (g(2$^+_1$)=+0.308) goes in the direction of the measured
g factors for $^{38}$Ar and $^{39}$K, the former being +0.24(12)
~\cite{args1} and the latter being +0.26~\cite{exp3}, indicating the
importance of configuration mixing. In addition, in the large space
there are further indications of substantial configuration mixing from
the considerable differences in the g(2$^+_1$) values between
$^{38}$Ar (g(2$^+_1$)=+0.308) and $^{46}$Ar (g(2$^+_1$)=+0.100) and
also between $^{40}$Ar (g(2$^+_1$)-0.197) and $^{44}$Ar
(g(2$^+_1$)=-0.022). For the states of higher angular momentum, J= 4,
6 and 8, there is also a large amount of quenching of the g factors as
one goes from the small space to the large space. For example for the
unique configuration J=8 (J$_p$=2, J$_n$=6) in $^{40}$Ar and
$^{44}$Ar, the g factor changes from -0.389 to -0.195 in $^{40}$Ar and
from -0.389 to -0.150 in $^{44}$Ar. The measured value of g(2$^+_1$)
in $^{40}$Ar is very small, -0.02(2).  This small value was explained
in~\cite{args2}.\\

c. \underline{The B(E2) values in the small and large spaces}\\

The experimental B(E2) values (taken from ~\cite{exp2} and for
$^{40}$Ar from ~\cite{args6}), and the calculated B(E2) values in the
small and large spaces, are given in Table ~\ref{tab:be2}.

In the small space both the B(E2;$0_1^+ \rightarrow 2_1^+$) and the
B(E2;$2_1^+ \rightarrow 4_1^+$) each change by less than 16$\%$ as one
moves from $^{38}$Ar to $^{46}$Ar. In the large space, however, these
two calculated B(E2) values increase by respective factors of three
and five, being largest for $^{46}$Ar and $^{44}$Ar respectively.  The
above large increase in the calculated values indicates that in the
large space as one moves from $^{38}$Ar to $^{46}$Ar there is
greater collectivity, as well as a reduction in the role played by
the closure of the f$_{7/2}$ neutron shell at N=28, as configuration
mixing effects become more important.

The experimental values for B(E2; $0_1^+ \rightarrow 2_1^+$) increase
monotonically from $^{38}$Ar to mid-shell, at $^{42}$Ar, and then
decrease until $^{46}$Ar.  The observed values are closer to the small
space calculated values for $^{38}$Ar and $^{46}$Ar but closer to the
large space values for $^{40,42,44}$Ar.

The very small value of the B(E2:$0_1^+ \rightarrow 2_2^+$) in
$^{42}$Ar in the small space is due to seniority considerations; this
is discussed in Appendix A, where it is shown to be zero. In the 0
$\hbar \omega$ calculations the values obtained for this B(E2) are
49.56 and 24 $e^2fm^4$ with the WBT and SDPF-U interactions,
respectively.  The experimental value from ~\cite{lasty} is 19.5
$e^2fm^4$.

\section{Comparison of $0 \hbar \omega$ calculations using the WBT and SDPF-U interactions}

The SDPF interaction~\cite{oldsdpf} is newer than the WBT
interaction~\cite{wbt} and in addition its form was recently recast
slightly to handle nuclei with Z$>14$ differently than nuclei with
Z$\leq$14~\cite{newsdpf}. This interaction is denoted by SDPF-U.  The
Z$>14$ version of SDPF-U is the one used in this paper.

\begin{figure*}
\includegraphics[scale=0.8]{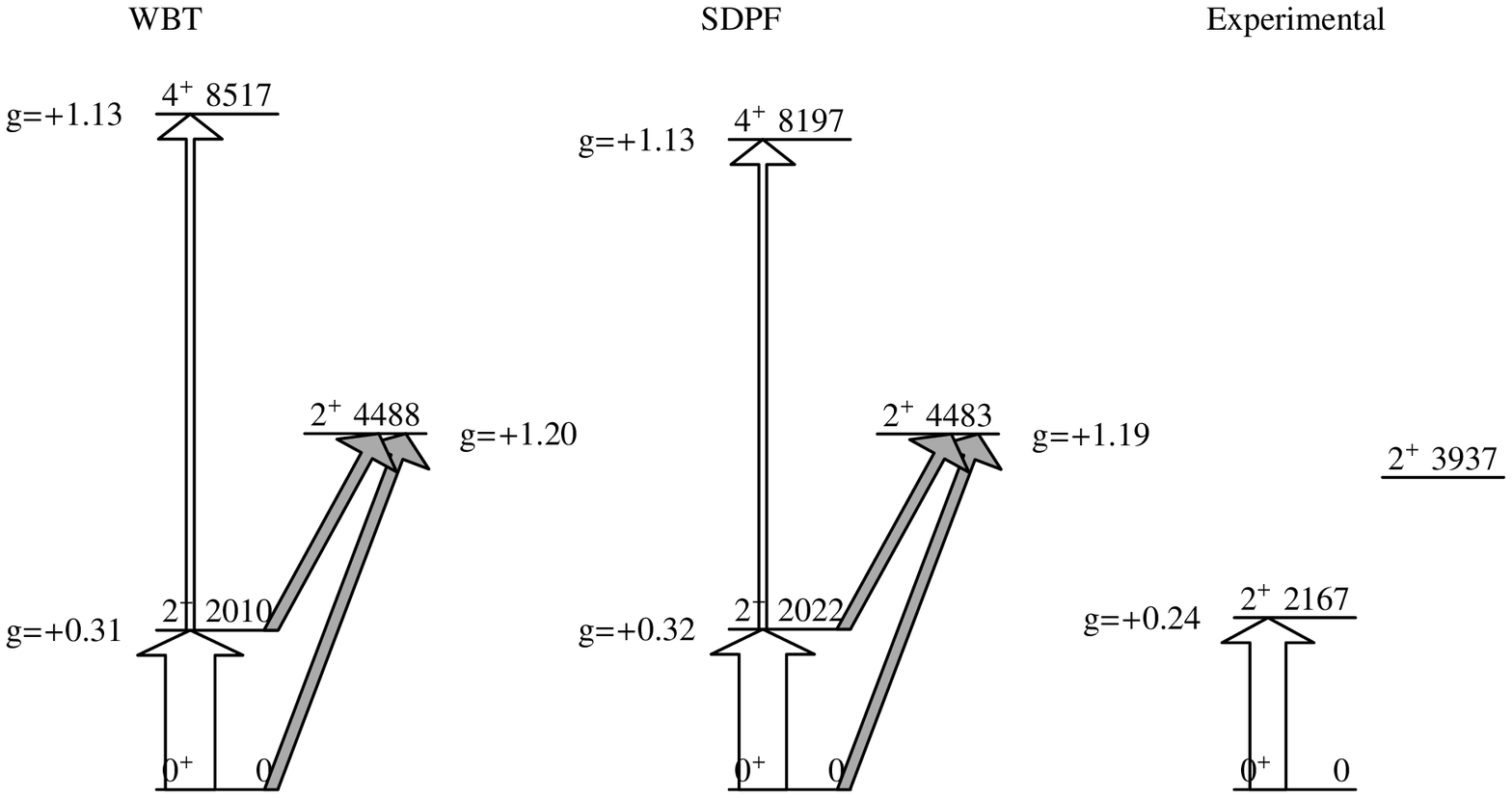}
\caption{$^{38}$Ar, levels are labeled by excitation energy in keV and g
  factor. B(E2 up) is proportional to width of the arrow.
\label{fig:38ar}}
\end{figure*}

\begin{figure*}
\includegraphics[scale=0.8]{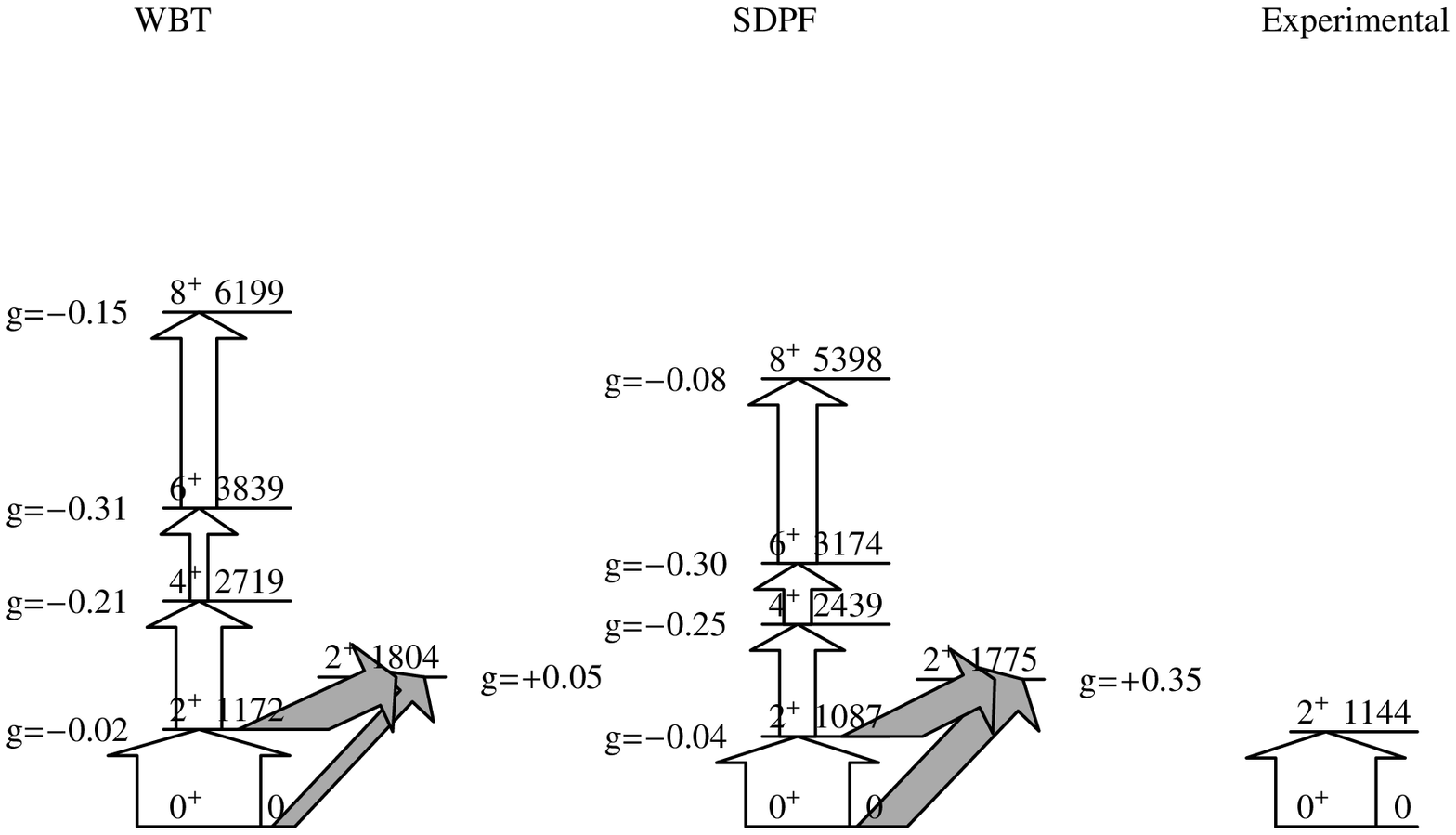}
\caption{$^{44}$Ar, levels are labeled by excitation energy in keV and g
  factor. B(E2 up) is proportional to width of the arrow.
\label{fig:44ar}}
\end{figure*}

\begin{figure*}
\includegraphics[scale=0.8]{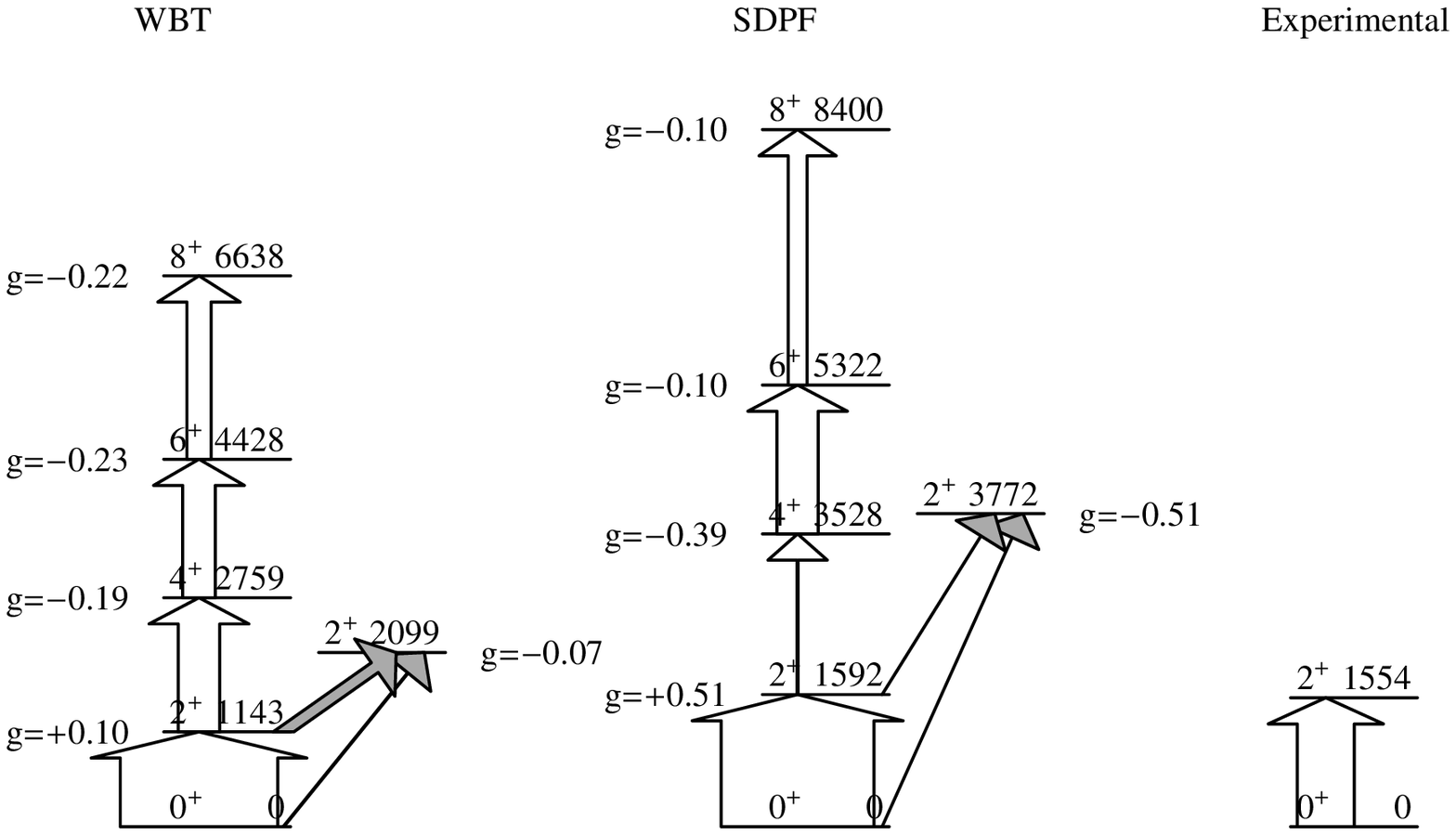}
\caption{$^{46}$Ar, levels are labeled by excitation energy in keV and g
  factor. B(E2 up) is proportional to width of the arrow.
\label{fig:46ar}}
\end{figure*}

As much previous work has been carried out using the WBT interaction,
it is valuable to compare the WBT results and conclusions to those
which the SDPF-U interaction offers.  The results for each nucleus are
presented in Figures ~\ref{fig:38ar} thru ~\ref{fig:46ar}, where the
excitation energies are given in keV. The numerical results are
simultaneously summarized for all five nuclei in Tables
~\ref{tab:ee},~\ref{tab:g}, and ~\ref{tab:be2}.  It is clear that, for
the lower mass even-even Ar nuclei, the computed nuclear structure
appears largely unchanged.  In $^{44}$Ar (Figure~\ref{fig:44ar}) some
distinctions between the results with the two interactions begin to
appear; in particular, the energies of the excited states above the
$2_1^+$ state are seen to diverge.

However, in $^{46}$Ar (Figure~\ref{fig:46ar}) the two interactions now
paint very different pictures of the low-energy nuclear structure of
this nucleus.  As was seen in~\cite{4546ar,n28}, the SDPF interaction,
before its recent alterations, did very well in accounting for those
energies of the levels of $^{46}$Ar that could be firmly
established. While the energies and g factors in $^{46}$Ar are showing
divergent behaviors between the WBT and SDPF-U interactions, we have
nearly identical $B(E2; 0_1 \rightarrow 2_1)$ values in this nucleus
for the two interactions, 541 in WBT and 525 $e^2 fm^4$ in
SDPF-U. However, these very similar calculated values for this B(E2)
differ considerably from the experimental value of 196 $e^2 fm^4$
reported in ~\cite{brown}.  

Indeed, in ~\cite{brown} the authors performed shell model
calculations for several nuclei including $^{46}$Ar. For that nucleus
they also obtained much too large a B(E2) value. They used the
Wildenthal interaction for the sd shell, the FPD6 interaction for the
fp shell, and the cross-shell interaction of Warburton, Becker,
Millener and Brown--see ref~\cite{brown} for further details.  On the
other hand, they got good agreement for the energy of the 2$^{+}_1$
state.  The fact that we here also obtain too large a B(E2) with yet
two other different interactions indicates that this is a robust
result.  However, as noted again in ~\cite{brown}, a different
approach, the self-consistent mean-field calculations by Werner et
al.~\cite{werner}, yields a much smaller B(E2) in close agreement with
experiment.  Using smaller effective charges will also reduce the
calculated B(E2) values.

One main thrust of our work is to see the effects of the various shell
model interactions on magnetic moments. Perhaps the most surprising
result is that, although for most of the Ar isotopes the g(2$_1^+$)
values with the SDPF-U are not so different from those of WBT, there
is a very large difference for $^{46}$Ar.  There, whereas for WBT the
value of g(2$_1^+$) is 0.100, for SDPF-U it increases to +0.513.
Likewise, the g(2$_2^+$) values in $^{46}$Ar are quite different,
-0.070 and -0.514, respectively. Furthermore, for the g(4$_1^+$) the
corresponding values are -0.190 and -0.388, respectively. These large
differences indicate the crucial importance of experiments to measure
these g factors in $^{46}$Ar in order to determine the efficacy of the
otherwise excellent SDPF-U interaction.

In order to understand the above noted differences between the results
of the two interactions, we look at selected configurations in
$^{46}$Ar in which the eight neutrons remain in the f$_{7/2}$ shell
and close that shell. Three of these configurations are displayed in
Table ~\ref{tab:occu}. We then focus on the proton configurations,
which are the only ones that matter in those cases.  The g factor for
the lowest single-particle energy configuration d$_{3/2}^2$
s$_{1/2}^2$ is small but the other two configurations have large
positive g factors.  What is most surprising is the very large
occupation probability of the d$_{3/2}^3$ s$_{1/2}$ configuration
(having a very large g factor, see Table~\ref{tab:occu}) in the
2$^+_1$ wavefunction with SDPF-U --21.77$\%$. This occupation
probability value is much larger than the corresponding result for the
WBT interaction where it is only 2.51$\%$.  This goes a long way in
explaining why the value of g(2$^+_1$) in $^{46}$Ar is larger for
SDPF-U than for WBT.

To explain the large percentage occupancy of the d$_{3/2}^3$ s$_{1/2}$
configuration for SDPF-U, we show in Table~\ref{tab:six} the observed
J=$\frac{3}{2}^+$ J=$\frac{1}{2}^+$ splittings in the odd K
isotopes. We see experimentally a steady reduction in this splitting
as one goes from $^{39}$K to $^{45}$K and a crossover in
$^{47}$K, where J=$\frac{1}{2}^+$ becomes the ground state.  In the
calculated $J=\frac{3}{2}^+$ - $J=\frac{1}{2}^+$ splittings in
Table~\ref{tab:six}, the WBT interaction gets some of the systematics
correct, narrowing the gap as we approach $^{47}$K and expanding it
afterwards. However the SDPF-U is much better in this region,
correctly giving the cross-over at $^{47}$K and returning the states
to the usual ordering at $^{49}$K.

The idea, as pointed out in~\cite{newsdpf}, is that the 0f$_{7/2}$
neutron - 0d$_{3/2}$ proton interaction is working to lower the
0d$_{3/2}$ orbital.  This makes the occupation of that orbital
increasingly favorable as we fill the 0f$_{7/2}$ orbit with neutrons.
This effect is considerably more pronounced in the SDPF-U interaction
than in the WBT interaction.  In our case, we note in Table
~\ref{tab:seven} the increasing removal of protons from the 1s$_{1/2}$
and their subsequent migration to the 0d$_{3/2}$ as the neutron number
increases from N=20 until N=28.  If neutrons are added past N=28, the
1p$_{3/2}$ neutrons behave in the opposite manner, working to make the
1s$_{1/2}$ orbit lower. (We do wish to point out that work on the
J=$\frac{3}{2}^+$ - J=$\frac{1}{2}^+$ splittings was carried out in more
limited shell model spaces some time ago by
Johnstone~\cite{lateadd}.)

There have also been discussions in the literature of the changes in
the neutron single particle energies~\cite{lastmin}. From the reaction
$^{46}$Ar(d,p) $^{47}$Ar these authors find reductions in both the f
and the p single particle spin-orbit splittings in $^{47}$Ar relative to its
isotone $^{49}$Ca. They attribute these changes to effects of the
proton-neutron tensor interaction in the former case and to the density
dependence of the spin-orbit interaction in the latter case.

It is interesting to calculate at N=28 the shell gap, i.e. the
p3/2-f7/2 neutron single particle energy difference.  This gap is
given by the binding energy difference expression BE(49Ca)+BE(47Ca) -
2 BE(48Ca). With the SDPF-U interaction we calculate a value of 4.74
MeV for this gap. This value can be compared to the corresponding gap
value from the experimental mass tables where we receive a value of
4.80 MeV in good agreement with the calculated value. We also note
that these values are both larger than the splitting of the
J=$\frac{3}{2}^-$ - J= $\frac{7}{2}^-$ splitting in $^{41}$Ca which is
1.9 MeV.

It should be emphasized that in the shell model calculations we
perform it is not necessary to put in by hand the changes in single
particle energies for different nuclei.  A good effective interaction
implicitly generates these changes, both for protons in the s-d shell
and for neutrons in the f-p shell.

\section{Conclusion}

Several interesting results were pointed out in this study. The
peculiar situation for $^{40}$Ar and $^{44}$Ar presents an excellent
chance to experimentally examine configuration mixing, since any
divergence in the experimentally measured properties of these two
nuclei represents the effects of configuration mixing. In the naive
shell model, they will have identical nuclear structure due to the
half filling of the d$_{3/2}$ orbital. The results of allowing 0
$\hbar \omega $ configuration mixing in the heavier isotopes of Argon
is examined. As we increase in mass to $^{46}$Ar we see a divergence
in the results obtained with the widely-used WBT interaction and the
newer SDPF-U interaction.  The low energy near-yrast nuclear structure
of $^{46}$Ar that is obtained with these two interactions is presented
and awaits more detailed experimental study.

\section{Appendix A}

The vanishing of the B(E2: $0^+_1 \rightarrow 2^+_2$) in a small space
calculation of $^{42}$Ar can be explained by noting that both the
proton and neutron configurations are at mid-shell d$_{3/2}^2$
f$_{7/2}^4$. The protons couple to J$_p$ the neutrons couple to J$_n$,
and J$_p$ and J$_n$ couple to J.

Here are all the configurations in the small space of [J$_p$ J$_n$]
which lead to a total J=0 or total J=2.

J=0           [0,0], [2,2]  [2,2']

J=2         [0,2]  [0,2']  [2,0]  [2,2], [2,2']  [2,4]   [2,4']

In the above J= 2' and J= 4' designate seniority v=4 states of the
f$_{7/2}^4$ neutron configuration.  The states J = 2 and J=4 have v=2
and J=0 has v=0.

Because we are at mid-shell the quantity s= (-1)$^{(v_p+v_n)/2}$ is a
good quantum number.  The wave functions have either s=+1 or s=-1.

     \underline{For J=0}

      s=1   is a linear combination (lc) of [0,0] and [2,2]

      s=-1 [2,2']

      \underline{For J=2}

      s=-1  is a lc of  [0,2] ,[2,0], [2,2'],[2,4']

      s=1   is a lc of  [0,2'],[2,2],[2,4]

At mid-shell for particles of one kind there cannot be a B(E2) between
states of the same seniority (see e.g. Lawson~\cite{lawson}).  We can
use this to show that the B(E2) from the state J=2 s=1 to J=0 s=1 must
vanish.

For the [2,2] to [2,2] transition we can break things down so that we
have a term with a transition from 2 to 2 for protons and a term with
a transition from 2 to 2 for neutrons. In both cases we get a v=2 to
v=2 transition which must vanish. The same story holds for the [2,4]
to [2,2] transition.  Consider next the [0,2'] term. It cannot connect
to [2,2].  Less obvious is the connection of [0,2'] to [0,0].This
vanishes because the one body E2 operator cannot change the seniority
by more than 2 units but here we require a change of four.  (See
eq. (A3.31) in Lawson~\cite{lawson}).

We have thus explained the vanishing B(E2) in $^{42}$Ar in the small space.

\begin{table}
\begin{center}
\caption{ Excitation energies in MeV for the even-even Argon isotopes}
\label{tab:ee}
\begin{tabular}{cccccc}
Excitation energies   &   $^{38}$Ar &$^{40}$Ar & $^{42}$Ar & $^{44}$Ar & $^{46}$Ar \\
&&&&&\\
\textbf{E(2$\mathbf{^+_1}$)}            &             &          &          &           &          \\
experiment            & 2.167 &1.464 &1.208  &1.144(17)& 1.550(10) \\
small space WBT       & 2.118 &1.254 &1.272  &1.254 & 2.118 \\
0 $\hbar \omega$ WBT  & 2.010 &1.424 &1.292  &1.172 & 1.143 \\
0 $\hbar \omega$ SDPF-U& 2.022 &1.281 &1.154  &1.087 & 1.592 \\
&&&&&\\
\textbf{E(2$\mathbf{^+_2}$)}            &             &          &          &           &          \\
experiment            &  3.937&2.524 &2.487  &  & \\
small space WBT       &  N/A & 3.642 & 2.926 &3.642 & N/A \\
0 $\hbar \omega$ WBT  &  4.488&2.865 & 2.327 &1.804 & 2.099\\
0 $\hbar \omega$ SDPF-U&  4.483&2.911 & 2.276 &1.775 &3.772 \\
&&&&&\\
\textbf{E(4$\mathbf{^+_1}$)}            &             &          &          &           &          \\
experiment            &     & 2.892  &   &  & \\
small space WBT       &  N/A & 2.750 & 2.422 &2.751 & N/A \\
0 $\hbar \omega$ WBT  & 8.517& 2.742 & 2.430 &2.719 & 2.759\\
0 $\hbar \omega$ SDPF-U&  8.197 &2.647& 2.165 &2.439 & 3.528\\
&&&&&\\
\textbf{E(6$\mathbf{^+_1}$)}            &             &          &          &           &          \\
experiment            &     &    &   &  & \\
small space WBT       & N/A & 3.670 &3.608 &3.670 & N/A \\
0 $\hbar \omega$ WBT  & N/A & 3.543 &3.731 &3.839 & 4.428 \\
0 $\hbar \omega$ SDPF-U& N/A & 3.204 &3.323 &3.174 & 5.322\\
&&&&&\\
\textbf{E(8$\mathbf{^+_1}$)}            &             &          &          &           &          \\
experiment            &     &    &   &  & \\
small space WBT       &  N/A &6.703  & 5.846 &6.703 & N/A \\
0 $\hbar \omega$ WBT  &  N/A &6.573 & 5.821 &6.199 & 6.638\\
0 $\hbar \omega$ SDPF-U& N/A & 6.308 & 5.302 &5.398 & 8.400 \\

\end{tabular}
\end{center}
\end{table}

\begin{table}
\begin{center}
\caption{  g factors in the even Argon isotopes}
\label{tab:g}
\begin{tabular}{cccccc}
g factors   &   $^{38}$Ar &$^{40}$Ar & $^{42}$Ar & $^{44}$Ar & $^{46}$Ar \\
&&&&&\\
\textbf{g($\mathbf{2_1^+}$)} &             &          &          &           &          \\
experiment            &  0.24(12)   &-0.02(2)    &   &  & \\
small space WBT       & 0.083  & -0.441  &-0.455  &-0.441 & 0.083 \\
0 $\hbar \omega$ WBT  & 0.308 & -0.197  &-0.095  &-0.022 &+0.100 \\
0 $\hbar \omega$ SDPF-U& 0.319 & -0.228 &-0.084  &-0.040 &0.513 \\
&&&&&\\
\textbf{g($\mathbf{2_2^+}$)} &             &          &          &           &          \\
experiment            &     &    &   &  & \\
small space WBT       & N/A  &-0.046  & -0.481 &-0.046 & N/A \\
0 $\hbar \omega$ WBT  & 1.198 &0.120  & 0.096 &0.045  &-0.070 \\
0 $\hbar \omega$ SDPF-U& 1.187 &0.136 &0.075 & 0.346 &-0.514 \\
&&&&&\\
\textbf{g($\mathbf{4_1^+}$)} &             &          &          &           &          \\
experiment            &     &    &   &  & \\
small space WBT       & N/A & -0.490 &-0.509 &-0.490 & N/A \\
0 $\hbar \omega$ WBT  & 1.134 &-0.354&-0.277 &-0.206 &-0.190 \\
0 $\hbar \omega$ SDPF-U& 1.132 &-0.357 &-0.289 &-0.246  &-0.388 \\
&&&&&\\
\textbf{g($\mathbf{6_1^+}$)} &             &          &          &           &          \\
experiment            &     &    &   &  & \\
small space WBT       & N/A &-0.525 &-0.515 &-0.525 & N/A \\
0 $\hbar \omega$ WBT  & N/A &-0.381 &-0.333 &-0.313 &-0.233 \\
0 $\hbar \omega$ SDPF-U& N/A & -0.394 &-0.328 &-0.301 &-0.095 \\
&&&&&\\
\textbf{g($\mathbf{8_1^+}$)} &             &          &          &           &          \\
experiment            &     &    &   &  & \\
small space WBT       & N/A & -0.389 &-0.477 &-0.389 & N/A \\
0 $\hbar \omega$ WBT  & N/A & -0.195 &-0.258 &-0.150 &-0.224 \\
0 $\hbar \omega$ SDPF-U& N/A & -0.188 &-0.255 &-0.084 &-0.095 \\         
\end{tabular}
\end{center}
\end{table}

\begin{table}
\begin{center}
\caption{ B(E2) values in $e^2 fm^4$ the even Argon isotopes}
\label{tab:be2}
\begin{tabular}{cccccc}
B(E2) values   &   $^{38}$Ar &$^{40}$Ar & $^{42}$Ar & $^{44}$Ar & $^{46}$Ar \\
&&&&&\\
\textbf{B(E2; $\mathbf{0_1^+ \rightarrow 2_1^+}$)} &             &          &          &           &          \\
experiment            &130(10)& 330(40) &430(100)  & 345(41) & 196(39) \\
small space WBT       &126.3 & 128.0 &146.5&128.6 & 134.2 \\
0 $\hbar \omega$ WBT  &178.2 & 251.9 & 338.4 &425.3 &541 \\
0 $\hbar \omega$ SDPF-U& 171 & 243 &351  &357 &525 \\
&&&&&\\
\textbf{B(E2; $\mathbf{0_1^+ \rightarrow 2_2^+}$)} &             &          &          &           &          \\
experiment            &     & 24(4)   & 20 (9)  &  & \\
small space WBT       &N/A  & 51.71 & 0.000 & 52.05 & N/A \\
0 $\hbar \omega$ WBT  &45.48 &41.3 &49.56 &57.82 & 2.081 \\
0 $\hbar \omega$ SDPF-U&43 & 51.8   &24   &125  &0.004 \\
&&&&&\\
\textbf{B(E2; $\mathbf{2_1^+ \rightarrow 2_2^+}$)} &             &          &          &           &          \\
experiment            &     & 49(26)   &   &  & \\
small space WBT       & N/A & 4.785 & 33.82 &4.815 & N/A \\
0 $\hbar \omega$ WBT  & 45.01 & 24.5 & 103.2  &130.8  & 40.71 \\
0 $\hbar \omega$ SDPF-U& 43    &17.79  & 98.3  &125.5  &0.221 \\
&&&&&\\
\textbf{B(E2; $\mathbf{2_1^+ \rightarrow 4_1^+}$)} &             &          &          &           &          \\
experiment            &     & 94(14)   &   &  & \\
small space WBT       & N/A & 48.10 & 52.46 &48.38 & N/A \\
0 $\hbar \omega$ WBT  & 28.97  & 70.14 & 59.64 &157 &143 \\
0 $\hbar \omega$ SDPF-U& 28 &78.77  & 75  &122.8  & 3.7 \\
&&&&&\\
\textbf{B(E2; $\mathbf{4_1^+ \rightarrow 6_1^+}$)} &             &          &          &           &          \\
experiment            &     &    &   &  & \\
small space WBT       & N/A & 18.18& 38.86  &18.26  & N/A \\
0 $\hbar \omega$ WBT  & N/A & 27.25  &37.02 &61.56 &112.6 \\
0 $\hbar \omega$ SDPF-U& N/A  &20.94  &65   & 95.187& 142.9\\
&&&&&\\
\textbf{B(E2; $\mathbf{6_1^+ \rightarrow 8_1^+}$)} &             &          &          &           &          \\
experiment            &     &    &   &  & \\
small space WBT       & N/A &33.33&44.65 &33.50 &N/A \\
0 $\hbar \omega$ WBT  & N/A &40.87 &80.99 &122.8 &82.91 \\
0 $\hbar \omega$ SDPF-U& N/A &35.5  &96   &134  & 64.7\\

\end{tabular}
\end{center}
\end{table}

\begin{table}
\begin{center}

\caption{Percentage occupancy and g factors of configurations in the
  the $2_1^+$ wavefunction of $^{46}$Ar in which the neutrons close
  the f$_{7/2}$ shell.}
\label{tab:occu}
\begin{tabular}{ccccc}
   &   &   &   &   \\
Proton configuration & Neutron configuration & SDPF-U & WBT & g factor\\
  d$_{3/2}^2$ s$_{1/2}^2$& f$_{7/2}^8$  & 1.08$\%$  &   1.96$\%$  & +0.0828\\
   d$_{3/2}^3$ s$_{1/2}$& f$_{7/2}^8$ & 21.77$\%$  &  2.51$\%$ &  +1.458\\
  d$_{5/2}^5$ d$_{3/2}^4$ s$_{1/2}^1$& f$_{7/2}^8$ &  3.60$\%$   & $< 1.0 \%$ &  +1.306\\
\end{tabular}
\end{center}
\end{table}

\begin{table}
\begin{center}
\caption{ The J=$\frac{3}{2}^+$ - J=$\frac{1}{2}^+$ splittings of the
  odd K isotopes in MeV.}
\label{tab:six}
\begin{tabular}{cccc}
   &   Experimental & WBT & SDPF-U \\
 $^{41}$K&    0.980476    & 1.106    & 0.854     \\
 $^{43}$K&    0.5612      & 1.109    & 0.672     \\
 $^{45}$K&    0.4745      & 0.871    & 0.345    \\
 $^{47}$K&   -0.3600      & 0.507    & -0.320    \\
 $^{49}$K&   0.200         & 0.729    & 0.078 \\
\end{tabular}
\end{center}
\end{table}

\begin{table}
\begin{center}
\caption{The average proton occupation in the yrast 2$^+$ state of even
  Argon isotopes for the SDPF-U (WBT) interactions.}
\label{tab:seven}
\begin{tabular}{ccc}
Neutrons   &   1s$_{1/2}$ average occupation&0d$_{3/2}$ average occupation\\
20  &   1.94  (1.94)&      2.09 (2.085)\\
22  &   1.76  (1.84)&      2.31 (2.22)\\
24  &   1.65  (1.80)&      2.45 (2.29)\\
26  &   1.45  (1.73)&      2.69 (2.38)\\
28  &   1.28  (1.66)&      2.86 (2.45)\\
\end{tabular}
\end{center}
\end{table}

\begin{figure*}
\includegraphics[scale=0.8]{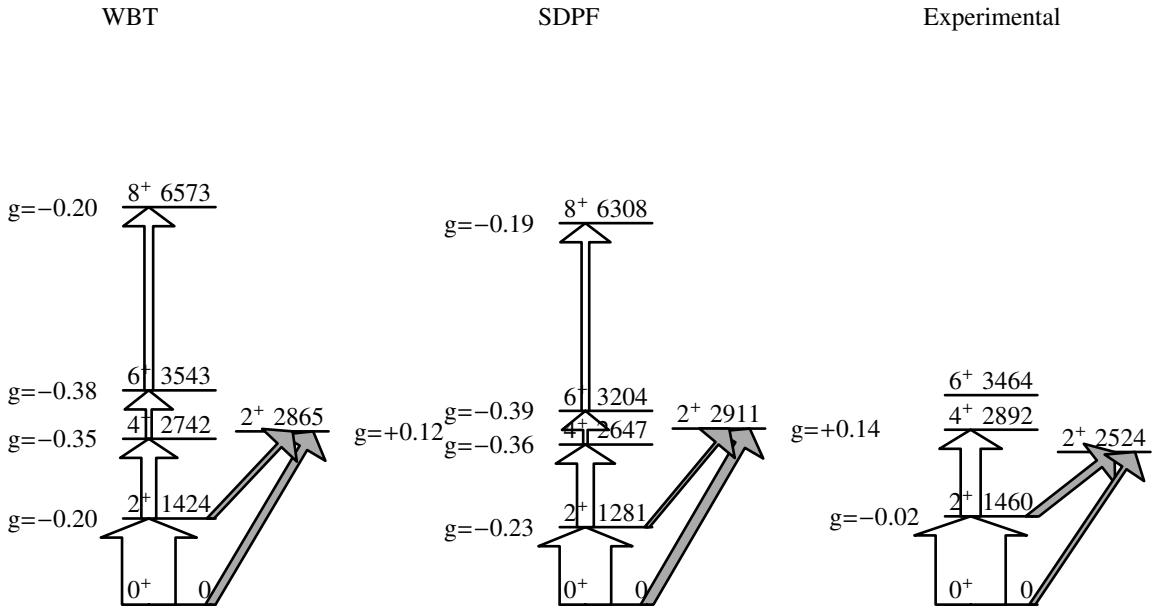}
\caption{$^{40}$Ar, levels are labeled by excitation energy in keV and g
  factor. B(E2 up) is proportional to width of the arrow.
\label{fig:40ar}}
\end{figure*}

\begin{figure*}
\includegraphics[scale=0.8]{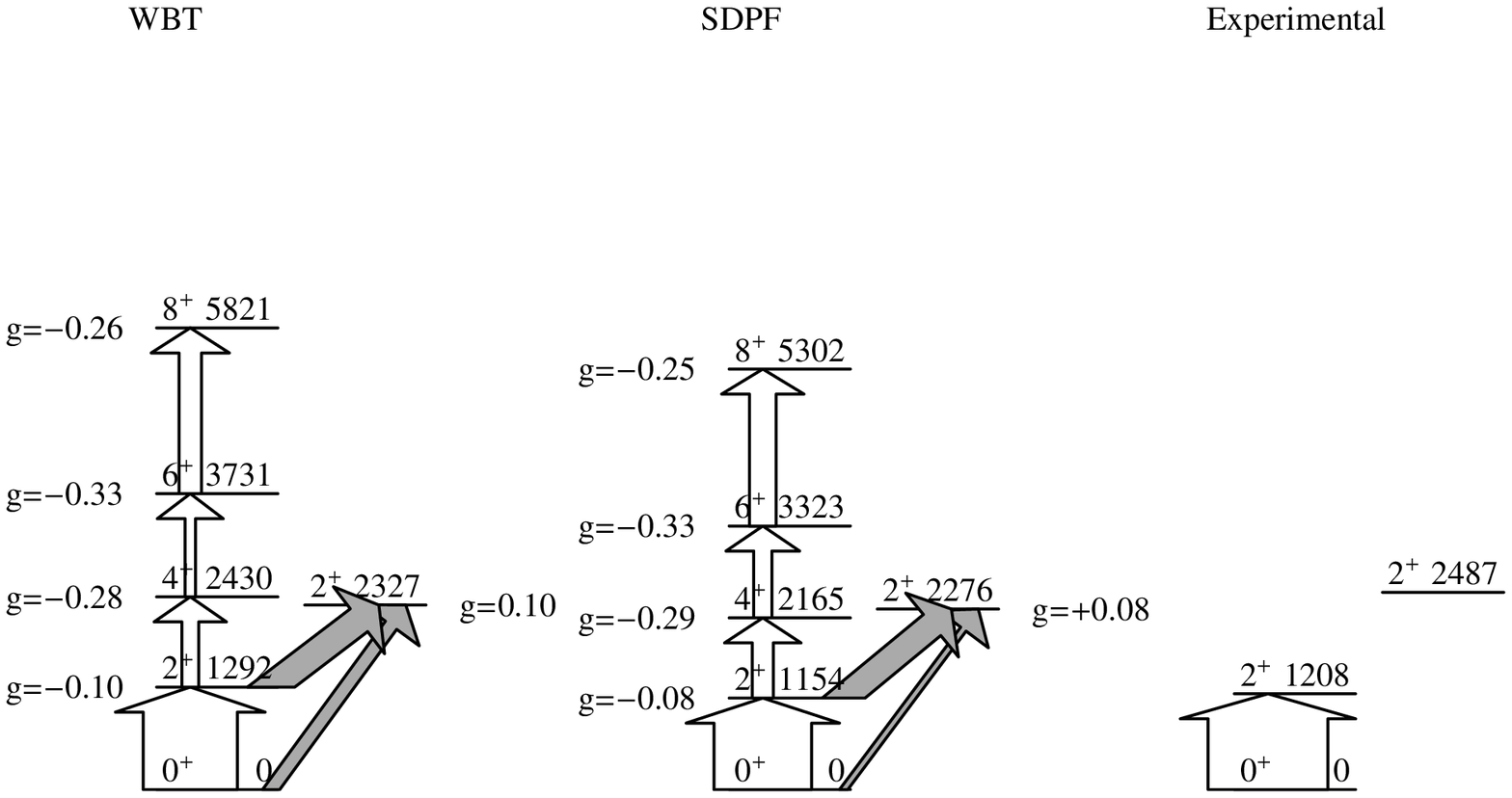}
\caption{$^{42}$Ar, levels are labeled by excitation energy in keV and g
  factor. B(E2 up) is proportional to width of the arrow.
\label{fig:42ar}}
\end{figure*}

ACKNOWLEDGEMENTS:\\ 

The authors would like to thank Gulhan Gurdal, Noemie Koller, and
Gerfried Kumbartski for their interest and encouragement.

\end{document}